\documentclass[sigconf]{acmart}
\renewcommand\footnotetextcopyrightpermission[1]{}
\usepackage{booktabs}
\usepackage{url}
\usepackage{xspace}
\usepackage{textcomp}
\usepackage{amsmath}
\usepackage{algorithm}
\usepackage[noend]{algpseudocode}
\usepackage{xspace}
\usepackage{balance}
\usepackage{array}
\usepackage{caption}
\usepackage{color}
\usepackage{subcaption}
\usepackage{multirow}
\usepackage{comment}
\usepackage{hyperref}
\usepackage{url}
\usepackage{makecell}
\usepackage{balance}
\setcopyright{none}

\begin{document}
\title{Leveraging Semantic and Lexical Matching to Improve the Recall of Document Retrieval Systems: A Hybrid Approach}

\author{Saar Kuzi}
\affiliation{\institution{University of Illinois at Urbana-Champaign}}
\email{skuzi2@illinois.edu}
\authornote{This work was done while interning at Google.}

\author{Mingyang Zhang}
\affiliation{\institution{Google Research}}
\email{mingyang@google.com}

\author{Cheng Li}
\affiliation{\institution{Google Research}}
\email{chgli@google.com}

\author{Michael Bendersky}
\affiliation{\institution{Google Research}}
\email{bemike@google.com}

\author{Marc Najork}
\affiliation{\institution{Google Research}}
\email{najork@google.com}

\fancyhead{}
\acmConference[arXiv Preprint]{}{}

\begin{abstract}
Search engines often follow a two-phase paradigm where in the first stage (the \emph{retrieval} stage) an initial set of documents is retrieved  and in the second stage (the  \emph{re-ranking} stage) the documents are re-ranked to obtain the final result list. 
While deep neural networks were shown to improve the performance of the re-ranking stage in previous works, there is little literature about using deep neural networks to improve the retrieval stage. 
In this paper, we study the merits of combining deep neural network models and lexical models for the \emph{retrieval} stage.
A hybrid approach, which leverages both semantic (deep neural network-based) and lexical (keyword matching-based) retrieval models, is proposed. 
We perform an empirical study, using a publicly available TREC collection, which demonstrates the effectiveness of our approach and sheds light on the different characteristics of the semantic approach, the lexical approach, and their combination.
\end{abstract}
\maketitle
\section{Introduction}
\label{sec:intro}
The ad hoc retrieval task is commonly addressed using a two-phase approach.
In the first stage (the retrieval stage), an initial result list of documents is retrieved from the collection for the query. 
Then, in the second stage (the re-ranking stage), the initial result list is re-ranked to generate the final list.
The focus of this work is on the retrieval stage where the main goal is to maximize the recall of the relevant documents retrieved. 
This is different than the goal of re-ranking which is to optimize the precision at high ranks of the final list.
Furthermore, since the retrieval stage is performed against all documents in the collection, a major requirement from a model is to be efficient.
The common practice for the retrieval stage is to use a lexical-based model, such as BM25 \cite{robertson1994some}. 
A lexical model assigns a relevance score to a document with respect to a query relying on the level of matching between the query and the document terms.
This type of model is likely to achieve a reasonable level of recall since the occurrence of the query words in documents is often a necessary condition for relevance. 
The lexical retrieval approach is also efficient due to the use of an inverted index.

A retrieval that relies only on a lexical model is likely to be non-optimal.
For example, such a model would have difficulty in retrieving relevant documents that have none of the query terms. 
This problem is partially a vocabulary mismatch problem in which a relevant document uses terms that are related to but different from the query terms.
Furthermore, relying solely on keyword matching may also not align well with people's actual information needs. When people search, what often they truly care about is whether the search results can address their needs, rather than whether the results contain the query words. 

To illustrate this point, an example query from our evaluation data set is presented in Table \ref{tab:doc-example}. In the table, we can see a passage from a relevant document retrieved using BM25 and a passage from a relevant document retrieved by the semantic model we used in this paper.
We can see that while the lexical document contains the query term ``fatality'', the semantic document contains a related term ``kill''. A further examination of the document revealed that the term ``fatality'' does not appear in any part. 
Thus, using a semantic model we can retrieve relevant documents that cover only some of the query terms.

The main idea of semantic matching of text is that it does not rely heavily on exact keyword matching. Instead, it measures complex relationships between words to capture semantics. Effective semantic models in recent years were mostly learned using deep neural networks \cite{devlin2018bert}.
Deep neural networks also attracted great interest in the IR community and many approaches for the re-ranking stage were devised \cite{onal2018neural}. The common main idea of the works on the subject is to use a large amount of training data, leveraging either query logs or weak supervision, 
to learn a model for the prediction of relevance between a document and a query. These works often follow the standard two-phase retrieval paradigm in which the retrieval stage is executed using a lexical-based model, and the result list is re-ranked using a neural network model.

The study of semantic models for the retrieval stage is a subject that was rarely studied in previous works. Two possible reasons for this can be: (1) semantic models tend to have lower recall due to their soft matching nature, and (2) before the recent development of fast approximate KNN search \cite{guo2020accelerating},
using neural networks for retrieval had a very high cost. This is because running a query through a neural model and pairing it with each of the documents in the collection is extremely inefficient.

In this work, we study the effectiveness of semantic models for the retrieval stage. 
Our main premise is that even if the recall of the semantic retrieval is low, it still can retrieve relevant documents not covered by the lexical model.
This is a reasonable assumption due to the complementary nature of the two approaches.
Thus, to benefit from both approaches, we propose a lexical-semantic hybrid retrieval approach.
The main idea is to run a semantic and lexical retrieval in parallel and merge the two result lists to create the initial list for re-ranking. Since the retrievals can be performed in parallel, our approach can be efficiently used in any system.

Besides the difference at which stage (retrieval vs. re-ranking) the model is used, another major difference between our model and many of the previously proposed neural models for IR~\cite{huang2013learning,xiong2017end,shen2014latent} is that our model does not require access to large-scale query logs. 
Inspired by the recent development of pre-trained language models~\cite{devlin2018bert}, we design weakly supervised learning tasks to learn corpus-specific semantics. This makes our model useful to learn domain-specific knowledge for a new search scenario and for systems where logs cannot be collected.

The suggested approach is deployment achievable for the following reasons: (1) the approach relies on adding a second retrieval source and is thus not expected to hurt the performance of the current lexical-based approach, (2) the neural model training is weakly supervised and no training data is in need, (3) an approximate KNN search is used which is very efficient and is not expected to affect the system latency, and (4) our method is fully implemented using open-source software and can be thus easily reproduced. 

An extensive empirical analysis of the proposed approach is performed using a public TREC collection. The analysis confirms that the semantic approach can retrieve a large number of relevant documents not covered by the lexical approach. 
Then, we show that by using a simple unsupervised approach for merging the result lists, significant improvements in the recall can be achieved.
Finally, an exploration of the different characteristics of the semantic and lexical retrieved documents is performed, using both quantitative and qualitative measures, that sheds light on the complementary nature of the two approaches.

To summarize, the main contributions of this work are: \begin{itemize}
    \item Proposing and studying a novel hybrid document retrieval approach that leverages lexical and semantic (neural network-based) models. The proposed approach is efficient enough to be deployed in any commercial system.
    \item Proposing an effective end-to-end weak supervision training approach for the retrieval stage that does not rely on any external resources.
    \item Conducting an empirical evaluation that demonstrates the effectiveness and robustness of the suggested approach compared to the lexical-only approach.
    \item Conducting an empirical study that illustrates the different characteristics of the lexical model, the semantic model, and their combination.
\end{itemize}

\begin{table}[h]
	\center
    \caption{\label{tab:doc-example} An example of relevant documents retrieved by the lexical and the semantic approaches. Only a part of the document which contains the relevant information is presented.}
	\begin{tabular}{l}
\toprule
\textbf{Query: ``Weather Related Fatalities''} \\
\hline
\makecell[l]{\textbf{Information Need}: A relevant document will report a type\\ of weather event which has directly caused at least one\\ fatality in some location.}\\
\hline
\textbf{Lexical Document} \\
\hline
\makecell[l]{``... Oklahoma and South Carolina each recorded three fatalities.\\
There were two each in Arizona, Kentucky, Missouri, Utah\\
and Virginia. Recording a single lightning death for the year\\
were Washington, D.C.; Kansas, Montana, North Dakota, ...''}
\\
\hline
\textbf{Semantic Document} \\
\hline
\makecell[l]{``... Closed roads and icy highways took their toll as at least\\
one motorist was killed in a 17-vehicle pileup in Idaho, a\\
tour bus crashed on an icy stretch of Sierra Nevada interstate\\
and 100-car string of accidents occurred near Seattle ...''}\\

\bottomrule
\end{tabular} 
\vspace{-10pt}
\end{table}

\section{Related Work}
\label{sec:related}

The main novelty of our work is that we study a lexical-semantic hybrid approach to improve the recall of the retrieval stage. While there has been a large body of work in the area of neural information retrieval (e.g., \cite{onal2018neural,shen2014latent,huang2013learning,guo2016deep,xiong2017end}), the focus was mainly on improving the re-ranking precision.

Semantic retrieval approaches that do not rely on deep neural networks were proposed in some previous works. 
In one line of works \cite{caid1995learned,atreya2011latent}, Latent Semantic Indexing (LSI) was used to generate dense representations for queries and documents which were either used alone for retrieval or combined with a lexical approach. 
The suggested approaches, however, demonstrated the limited ability of LSI in improving the effectiveness of the retrieval stage.
In another work \cite{boytsov2016off}, KNN search was used for semantic retrieval by leveraging a statistical translation model.
In this work, our focus is on studying neural network-based approaches.

There have been some previous works on developing neural network-based semantic approaches for the retrieval stage of documents.
One work \cite{zamani2018neural} proposed 
a model that learns sparse vectors for documents and queries which can be used for retrieval with an inverted index. In another work \cite{gysel2018neural}, KNN search was used for the retrieval stage with neural network-based embeddings. The suggested approach \cite{gysel2018neural}, however, is not applicable for large collections since it requires the learning of document-specific representations for the entire collection.
In our paper, the focus is on studying the integration of lexical and neural approaches in the general case. 
Thus, our approach can be applied on top of any semantic model to further improve its performance.
Furthermore, the approach we take in this paper uses an existing neural model with some small modifications, whereas in those previous works new models were designed for the task. For this reason, our approach can more easily leverage novel neural models in the future.

A lexical-semantic hybrid approach was previously studied for the re-ranking stage \cite{mitra2017learning}. Specifically, two neural networks were trained jointly accounting for local (term-based interactions) and distributed (semantic) representations of queries and documents. In this work, we show that a hybrid approach can also help to increase the recall of the retrieval stage.

The recent success of applying the pre-trained language model BERT \cite{devlin2018bert} to many NLP tasks motivated the development of several BERT-based re-ranking models for IR \cite{nogueira2019passage,nogueira2019multi,yilmaz2019cross}.
The main idea of these works is to treat the query and the document as two consecutive sentences in BERT and use feed-forward layers on top of BERT's classification layer to compute the relevance score.
This approach was used for re-ranking of passages \cite{nogueira2019passage,nogueira2019multi}, and more recently to re-rank news-wire documents \cite{yilmaz2019cross}. 
Motivated by the success of BERT for the re-ranking task, in this work, we use the BERT architecture for retrieval. Differently from previous works, we take a representation-based approach, by generating embedding vectors, which is more applicable for the retrieval stage.

Neural network-based semantic retrieval models were already applied to several other applications rather than document retrieval. In one work \cite{dai2019context}, BERT was used for weighting terms in the inverted index of passages.
In another work \cite{mitra2019incorporating}, an efficient neural re-ranking and retrieval approach was suggested by assuming independence between query terms. This approach \cite{mitra2019incorporating}, however, was mainly studied for the re-ranking of passages.
Finally, neural models were shown to be more effective than lexical models for the retrieval stage in QA systems, conversational agents, and product search \cite{aliannejadi2019asking,yang2017neural,lee2019latent,nigam2019semantic}.

Recall can also be improved through query expansion \cite{carpineto2012survey}.
This approach, however, is often not used in commercial systems due to efficiency issues. First, query expansion uses very long queries which result in a prohibitive query evaluation time \cite{lavrenko2006real,billerbeck2004techniques,theobald2005efficient}.
Second, the most effective approach, which relies on the result list to learn expansion terms (pseudo-relevance feedback) \cite{xu2017quary,abdul2004umass}, requires two sequential retrieval steps and is thus not efficient enough. 

Document expansion is another technique that is used for improving the recall of retrieval systems \cite{tao2006language}. Recent works have demonstrated the effectiveness of this approach for the retrieval of passages \cite{nogueira2019document,nogueira2019doc2query}. Using it for document retrieval, however, was shown to have limited effectiveness \cite{billerbeck2005document}.
\section{A Hybrid Retrieval Approach}
\label{sec:model}
\begin{figure}[t]
\center
\includegraphics[width=8cm,height=3.7cm]{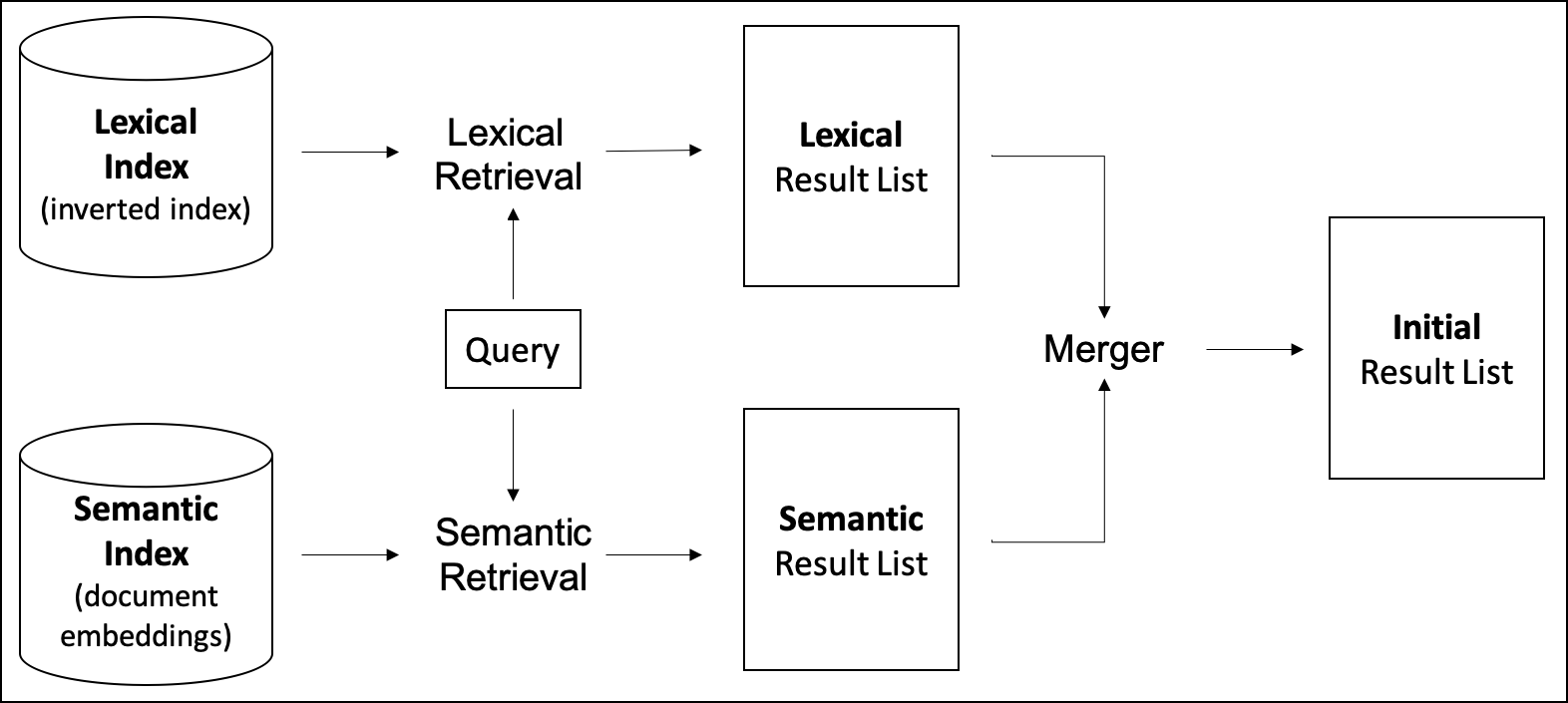}
\caption{\label{fig:framework} The hybrid retrieval approach.}
\end{figure}

In this paper, the focus is on the retrieval stage where the goal is to retrieve an initial set of documents of size $c$ using both semantic and lexical models.
The next step, which is out of the scope of this research, is the re-ranking stage in which the initial result list is ranked to generate a final list of size $c'$ (usually, $c' \ll c$).

The hybrid approach is depicted in Figure \ref{fig:framework}. The approach requires the existence of two indexes: (1) a lexical index (an inverted index), and (2) a semantic index (document embeddings matrix). Given a query $q$, two retrieval steps are performed in parallel. Lexical retrieval is performed in which the words in the query are matched with the words in documents. In this paper, we use the BM25 model \cite{robertson1994some} which is a classical retrieval approach that is highly effective and widely used by current retrieval systems. (For example, BM25 is the main approach taken by systems in recent IR competitions \cite{craswell2020overview}.) Semantic retrieval is also performed by first inferring an embedding vector for the query and then performing KNN search against the semantic index. The two result lists, each of size $c$, are pooled and then a merger is used to select $c$ documents from the pool to obtain the initial result list.

The hybrid retrieval approach was developed to be efficient enough so that it could be deployed in any system.
Our main goal is to avoid any extra overhead on top of the lexical (inverted index-based) approach which is the standard in current systems. The hybrid approach, by using two independent retrieval stages (semantic and lexical), can achieve this goal since the two can be performed in parallel. Furthermore, since we use approximate KNN search for the semantic retrieval \cite{muja2014flann,guo2016quantization,guoaccelerating}, it is expected to be as efficient as an inverted index-based search \cite{li2014two,boytsov2016off}.

In the remainder of this section, we cover the technical details regarding the implementation of the hybrid approach including details about the semantic retrieval implementation as well as the merging step.

\subsection{Semantic Retrieval}
This section describes the details of the neural model used for the semantic retrieval part. 
It is important to mention that in this work we are not interested in the full optimization of the semantic (neural) model but to study the potential benefits of combining semantic and lexical result lists. To that end, we make implementation decisions mainly in light of the findings of recent works on language understanding to obtain a sufficiently effective semantic model. 
Studying the effectiveness of different semantic models for the hybrid approach is left for future work.

\subsubsection{Neural Model Architecture}
\begin{figure}[t]
\center
\includegraphics[width=6.7cm,height=4.6cm]{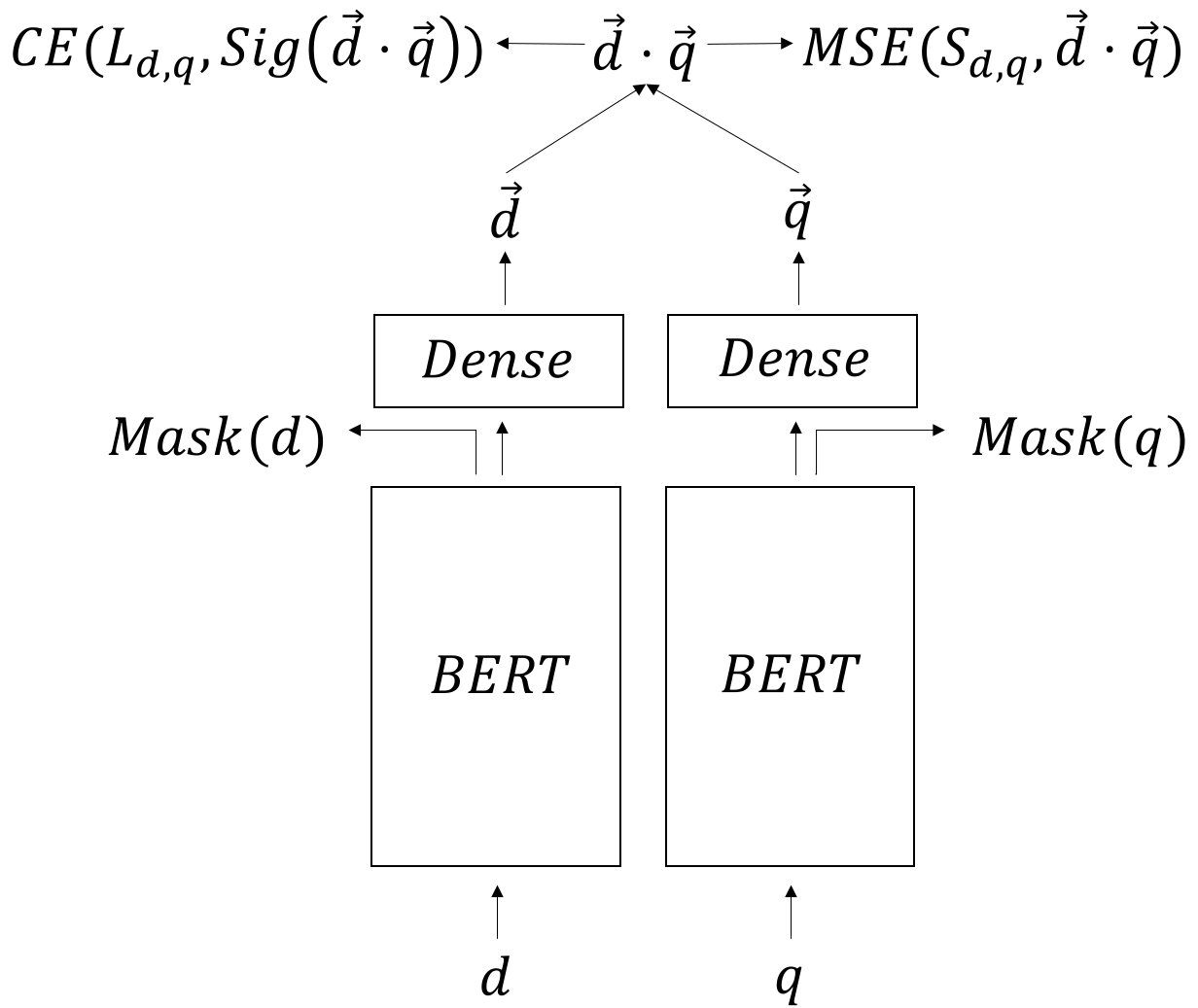}
\caption{\label{fig:arch} The neural network architecture of the semantic retrieval model.}
\end{figure}
The main idea of semantic retrieval is to generate query and document embedding vectors. Then, at serving time, a semantic similarity between a query and a document can be measured using the cosine function. 
The general architecture of the neural network, which was used for the semantic model, is depicted in Figure \ref{fig:arch}.
To generate query/document embeddings, we adopt the early idea of Siamese neural network architectures \cite{bromley1994signature}; this architecture was selected since it enables us to obtain query-independent document representations for indexing. 
Specifically, we are given a neural model that gets as an input a sequence of words and outputs a continuous vector. 
This model is used to generate both query and document vectors in parallel. 
In this paper, the architecture of the BERT model was used \cite{devlin2018bert}.
We chose this model as it was shown to achieve state-of-the-art performance in many NLP tasks. To generate an embedding vector for a document/query, we collect the pooled output from BERT and add an extra dense layer on top of it. The parameters of the BERT module are shared by the query and the document model to learn the common knowledge in the text. The parameters of the top dense layers of the query and the document model are trained separately so that we can learn query- and document-specific representation.
Then, the dot product between the vectors serves as a predicted \emph{relevance score} of the document to the query. The loss function for a pair of a query $q$ and a document $d$, which is associated with a binary relevance label $L_{d,q}$ and a continuous relevance score $S_{d,q}$, is defined as:
$$\mathcal{L} = \mathit{CE}(L_{d,q}, \mathit{Sigmoid}(\vec{d}\cdot \vec{q})) + \mathit{MSE}(S_{d,q}, \vec{d}\cdot \vec{q}) + \mathit{Mask}(q) + \mathit{Mask}(d).$$
Where $\mathit{CE}$ and $\mathit{MSE}$ are the Cross Entropy loss and Mean Squared Error loss, respectively; $\mathit{Mask}(\cdot)$ is the masked language model loss used in BERT; $\vec{q}$ and $\vec{d}$ are the vectors generated by the neural model.
We use the two losses as it is expected for the two to be complementary. While the $CE$ loss can help learn the rough distinction between something that is completely non-relevant to something that is somehow relevant. The $MSE$ loss can fine-tune the model to be more discriminative. We tried to fine-tune the model with just $CE$ and $MSE$ loss at the end of the training process but didn't notice much difference. Probably this is because differently from the original BERT paper, here we are directly training a model on the target data set. 

\subsubsection{Training data}
Semantic retrieval models, learned using deep neural networks, require large amounts of training data which is often hard to obtain.
To address this issue, several previous works have explored using weak supervision for the re-ranking task \cite{dehghani2017neural,nie2018multi,haddad2019learning}.
In this work, we also use weak supervision and demonstrate its effectiveness for the retrieval stage. 
Furthermore, unlike previous works, we propose an end-to-end training data generation pipeline that does not rely on any auxiliary resources.
Generalizing the results obtained in this work to semantic models that were learned using labeled data is an important direction worth exploring in future research (when such data is available).
Our proposed framework is general enough to facilitate the study of this direction.

To obtain training queries, tri-grams and bi-grams that appear in at least 5 documents in the collection are extracted.
Then, queries with less than 10 results when using BM25 are filtered out to make sure that we have enough training data for learning effective representations.
Next, document-query pairs, associated with a relevance score and a binary relevance label, are generated using a weak supervision approach (similarly to a previous work \cite{dehghani2017neural}).
For each query, 10 documents are retrieved using BM25 and each document is replaced by at most 5 passages from it.\footnote{A document is split into passages of 20 words with a sliding window of size 10.} Only passages that contain all query terms are used.
We use passages instead of using the entire document due to the limitation of BERT in handling long sequences of words \cite{dai2019transformer}. The query-document pairs, which are generated using our approach, are considered relevant.
Non-relevant pairs are generated using random sampling. 
To create relevance scores for query-document pairs, we randomly remove query terms from a relevant passage and replace them with random terms from the vocabulary.
Specifically, a pair of a bi-gram query and a relevant passage will be transformed into three pairs by adding two more pairs where the passage only matches a single term. To determine the match score, any relevance measure score like BM25 can be used. In practice, we found that using predefined scores works pretty well. That is, the full match score is set to $1$, while the partial matching score is set to $0.6$. 
Similarly, a pair of a tri-gram query and a relevant passage will be transformed into seven pairs; the full match score will be $1$, while the partial matching score will be $0.55$ and $0.65$ for single and double matching, respectively.

\subsubsection{Retrieval}
After the model was learned, it can be used to generate the semantic index by inferring vectors for all passages in the collection in an offline manner.
Then, at serving time, KNN search can be used for semantic retrieval. 
Since we have passage embeddings rather than document embeddings, there is a need to transform the result list to the document level. To do that, we sum up the scores of retrieved passages per document to obtain a document score.

\subsection{Hybrid Merging}
In this step, the documents retrieved by the semantic and the lexical approach are pooled to create a document set of size up to $2c$.
Then, a merger function assigns a score to every document in the pooled set.
Finally, $c$ documents with the highest scores are used to form the initial result list.

Using either the lexical or the semantic model as the merger function is likely to favor documents from only one of the two models. This is not desirable since we are interested in having both semantic-based and lexical-based relevant documents in the final list.
When using neural networks for re-ranking, previous works tended to rely on semantic scores because their retrieval stage has already enforced lexical matching (e.g., \cite{guo2016deep,dehghani2017neural}). 
For the retrieval stage, however, relying on semantic scores may not be the best choice.
One reason for this is that to generate semantic scores for the documents returned by the lexical approach, we need to run them against the neural model which may be inefficient. 
Furthermore, our preliminary examinations showed that the relevant documents in the semantic result list do not necessarily appear in high ranks. This suggests that semantic retrieval is not as discriminative as a lexical one.
This is probably because embeddings can be regarded as smoothed representations of text and are hence not discriminative enough.
On the one hand, they are strong at finding semantically similar text; on the other hand, facing a piece of semantically matched text and a piece of exactly matched text, as their smoothed representations would be quite similar, just relying on semantic representations to rank them may not be very effective. 

To address those issues, we use the relevance model RM3 \cite{abdul2004umass} as a merger which was shown to be an effective approach for TREC-style documents in some previous works (e.g., \cite{yang2019simple}). 
RM3 is essentially a probability distribution induced from the top documents in the initial result list and the original query which is supposed to serve as a representation of the user's information need; we refer the reader to the original paper \cite{abdul2004umass} for more details about this model.
RM3 is used as a merger in the following way. 
First, an RM3 model is induced from the result list of the lexical model (we use the lexical results since semantic scores are not as discriminative as lexical scores). 
Then, each document in the pooled set is scored using the RM3 model. Finally, the $c$ documents with the highest scores are selected to form the initial result list. 
Using RM3 is advantageous in this scenario because it takes into account the lexical similarity between the query and the document as well as the similarity between the document and related terms which can be indicative of semantic similarity.
We note that other approaches for the merger step can also be used. Yet, as will be shown in the experimental section, using RM3 already results in significant improvements and is simple and easy to implement.
From a practical point of view, it is important to mention that since we use RM3 only to score the documents in the pooled set, the query processing time is not supposed to increase largely. This is contrary to the common use of RM3 for pseudo-relevance feedback which requires two independent retrieval steps.
\section{Evaluation}
\label{sec:exp}

\subsection{Experimental Setup}
\subsubsection{Data set} A TREC collection (disks 1\&2) of 441,676 news-wire documents was used for the evaluation. The titles of TREC topics 51-200 served as queries.
This collection was selected since our focus is on performing a systematic analysis of retrieval models that rely solely on textual data. Thus, we are interested in a collection that has minimal noise and that contains reliable relevance judgments.
Since the focus in this work is on weak supervision-based semantic models, our method does not require large data sets of labeled data, and we thus leave the evaluation on such data sets (for example, the TREC DL data set \cite{craswell2020overview}) for future work.

Using this collection, our training data set ended up having 3.8M bi-gram queries, 1.7M tri-gram queries, and about 1B training examples (passage-query pairs).  
As already mentioned in the previous section, we split the documents in the collection into passages, resulting in approximately 22M passages. Thus, to generate an effective result list of documents, a large enough number of passages is needed to obtain enough evidence regarding each document. In this paper, we empirically set this value to 10,000.

\subsubsection{Lexical model implementation} 
BM25 was used as the lexical model (denoted \textbf{Lexical}).  
The Anserini toolkit \cite{yang2017anserini} was used for document and query pre-processing and for the implementation of the BM25 model (used as a baseline or as part of the hybrid approach) and of the RM3 model (used in the merging step of the hybrid approach). 
RM3 was not used as a baseline since it requires two consecutive retrieval steps and is thus not applicable to many search applications.
The free parameters of the lexical approaches were set to default values.\footnote{github.com/castorini/anserini}
One of the reasons for choosing Anserini is that its default free parameters for the lexical models are tuned to produce highly effective results for TREC collections \cite{yang2017anserini}. 
Krovetz stemming and stopword removal were applied to both queries and documents. For the evaluation, only queries for which all query terms are in the vocabulary of the semantic model were used (121 queries). We limited the evaluation to these queries to study the benefits of the lexical-semantic integration for queries that can potentially benefit from both. 
We thus leave the evaluation of other queries for future work.

\subsubsection{Semantic model implementation}
We do not use a pre-trained model; instead, we architected a BERT model using the TensorFlow library with 6 layers, a hidden size of 256, and 4 attention heads, and trained it using the Adam optimizer with a learning rate of 5e-4 and a batch size of 32 for 5 million training steps.
We use a vocabulary of 7500 words which was obtained by using a threshold of 300 occurrences of a word in the training set. 
The semantic retrieval was performed using an approximate in-memory KNN search to enable the efficient parallel execution of the semantic and the lexical retrieval.\footnote{github.com/google-research/google-research/tree/master/scann}\\

\subsubsection{Evaluation measures}
Since our focus is on improving the recall of retrieval, we report the following evaluation measures: $recall$, Mean Average Precision ($MAP$), and the total number of relevant documents retrieved for all queries ($\# rel$). Unless stated otherwise, those measures are calculated using the full size of the result list, $c$ ($\in \lbrace 500, 1000, 1500, 2000 \rbrace$).
To measure the robustness of the hybrid approach, we also report the Reliability of Improvement ($RI$). $RI=\frac{|Q^{+}|-|Q^{-}|}{|Q|}$, where $|Q^{+}|$ and $|Q^{-}|$ are the number of queries for which the hybrid approach performs better or worse than the lexical baseline, respectively; $|Q|$ is the total number of queries. The two-tailed paired t-test was used to determine statistically significant differences between different methods ($pval<0.05$). 

\subsection{Experimental Results}
\begin{table}[t]
	\center
    \caption{\label{tab:oracle} The potential improvements in terms of recall of the hybrid approach over the lexical approach. All differences with Lexical are statistically significant.}
	\begin{tabular}{ccccc}
\toprule

Method & $c=500$ & $c=1000$ &
$c=1500$ & $c=2000$ \\
\hline

Lexical & $.429$ & $.538$ & $.596$ & $.635$ \\ 
Semantic & $.063$ & $.106$ & $.137$ & $.163$ \\ 
Hybrid & $.454$ & $.568$ & $.628$ & $.669$ \\ 
\hline
\% Improvement & +5.8\%& +5.6\%&+5.4\% &+5.4\%\\
\bottomrule
\end{tabular}

\end{table}

\subsubsection{The potential benefits of the hybrid approach} As a first step, we are interested in examining the potential benefit of enriching a lexical-based result list using documents retrieved by a semantic model. Specifically, we are interested to know to what extent the semantic approach can retrieve documents that were not retrieved (or ranked low) by the lexical retrieval model. The results of this analysis can serve as an upper bound for the performance of the hybrid approach. 
To measure the potential benefits of the hybrid approach, the following experiment was performed. Given two result lists of size $c$ (lexical and semantic), a final result list of size $c$ is generated as well. 
To do that, we identify relevant documents in the semantic-based result list that do not appear in the lexical-based list.
Then, we replace the non-relevant documents in the lexical list with the semantic-based relevant documents.\footnote{Since we focus on recall, there is no importance for the order of replacement.} The results of this experiment are reported in Table \ref{tab:oracle}. 
According to the results, we can see that the lexical approach is much more effective than the semantic approach in terms of recall for all sizes of the result list. This result shows that the semantic approach cannot replace the classical lexical model in the retrieval stage and explains why previous works only used neural models for the re-ranking stage (e.g., \cite{dehghani2017neural,guo2016deep}).
Yet, this analysis reveals that a semantic model can retrieve a large number of relevant documents that are not included in the lexical-based result list. Specifically, for all sizes of the result list, there is a large and significant improvement in recall when incorporating semantically retrieved results in the lexical list. Furthermore, it is interesting to see that the improvement is stable with respect to the result list size which attests to the potential robustness of the hybrid approach. 
This result motivates the exploration of automatic approaches for merging the two lists. In the next sections, we show that even when using a simple unsupervised merging approach, significant improvements can be achieved. 

\subsubsection{Hybrid approach performance}

\begin{table}[t]
	\center
    \caption{\label{tab:main} The performance of the hybrid retrieval approach. All differences in performance ($MAP$ and $recall$) between methods in each block are statistically significant.}
	\begin{tabular}{cccccc}
\toprule
$c$ & Method & $recall$ & $MAP$ & $\# rel$ & $RI$ \\ 
\hline
500 & Lexical & $.429$ & $.225$ & $11,585$ & - \\ 
& Hybrid & $.441\:(+2.8\%)$ & $.228$ & $11,949\:(+3.1\%)$ & $.413$ \\
\hline
1000 & Lexical & $.538$ & $.256$ & $15,386$ & - \\ 
& Hybrid & $.553\:(+2.8\%)$ & $.259$ & $15,848\:(+3.0\%)$ & $.512$ \\
\hline
1500 & Lexical & $.596$ & $.269$ & $17,487$ & - \\ 
& Hybrid & $.612\:(+2.7\%)$ & $.272$ & $18,033\:(+3.1\%)$ & $.488$ \\
\hline
2000 & Lexical & $.635$ & $.275$ & $18,997$ & - \\ 
& Hybrid & $.653\:(+2.8\%)$ & $.278$ & $19,613\:(+3.2\%)$ & $.446$ \\
\bottomrule
\end{tabular}
 
\end{table}

The performance of the hybrid approach is reported in Table \ref{tab:main}.
The results demonstrate the effectiveness of the hybrid method even when a simple approach is used for the merging stage. Specifically, for all levels of $c$, the hybrid approach improves over the baseline lexical approach in terms of recall by about $3\%$. Focusing on the RI measure, we can see that the hybrid approach is also highly robust with respect to the different queries in terms of the recall improvements.

\begin{figure}[t]
\center
\includegraphics[width=6.5cm,height=4.8cm]{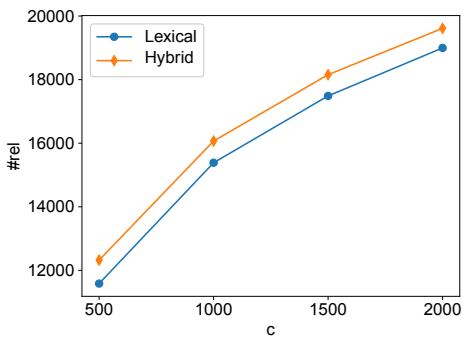}
\caption{\label{fig:len-analysis} The number of relevant documents when merging a fixed-length semantic-based result list (of 2000 documents) with a lexical-based result list of different lengths.}
\end{figure}

An important question that comes up from the results in Table \ref{tab:main} is: Can the same improvements in recall be achieved by simply considering a longer result list of the lexical model and re-ranking it using RM3? To address this question, the following analysis was performed. Focusing on a semantic-based result list of $2000$ documents, we merge it with lexical-based result lists of increasing lengths ($\in\left\lbrace{500, 1500, 1000, 2000}\right\rbrace$), and clip the final result lists to the original length of the lexical result list. 
The results of this analysis are presented in Figure \ref{fig:len-analysis}. In the figure, we report the number of relevant documents retrieved for each size of the result list. As can be seen, the number of relevant documents added by the hybrid approach remains stable for all lengths of the lexical list (the value is around 700). This analysis shows that even though we consider longer lexical lists, the semantic approach can still bring the same amount of unique relevant documents on top of it.

\subsubsection{Robustness analysis}

\begin{table}[b]
	\center
    \caption{\label{tab:quarters} The performance (recall) of four equally sized groups of queries, partitioned based on their performance when the lexical model is used. Statistically significant differences are marked with an asterisk.}
	\begin{tabular}{ccccc}
\toprule
Method & Q1 & Q2 & Q3 & Q4 \\ 
\hline
Lexical & $.167$ & $.423$ & $.663$ & $.887$ \\
Hybrid & $.191^{*}$ & $.446^{*}$  & $.674^{*}$ & $.891$ \\
\hline
\% Improvement & +14\% & +5.5\%& +1.7\%& +0.5\%\\
\bottomrule
\end{tabular} 
\end{table}

\begin{figure}[t]
\center
\scriptsize
\includegraphics[width=6.5cm,height=4.8cm]{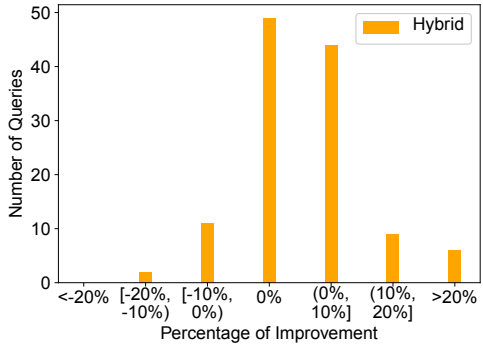}
\caption{\label{fig:hybrid-robust} The number of queries in different groups that were divided based on similar level of decrease/increase in performance in the hybrid approach as compared to the lexical retrieval model (in percentage).}
\end{figure}

In this section, we analyze the robustness of the hybrid approach with respect to the different queries.
First, we divide the queries in the evaluation set such that the queries in each group have a similar level of increase (or decrease) in recall when using the hybrid retrieval approach, compared to the lexical retrieval baseline; the increase/decrease is measured in percentage; we focus on a result list of 1000 documents. The queries in each group are counted and presented in a histogram in Figure \ref{fig:hybrid-robust}.
According to the results, it is clear that the hybrid approach is very robust. Specifically, the hybrid approach either improves or does not degrade the performance of the baseline in the majority of cases. According to the results in Figure \ref{fig:hybrid-robust}, for $50\%$ of the queries there is an improvement when using the hybrid approach, for $40\%$ there is no change in performance, and for $20\%$ there is a degradation in performance. 
Yet, while the average percentage of improvement for the good performing queries is around $18\%$, the performance of the bad performing ones decreases in about $4\%$ only. 

\begin{figure*}[t]
\center  
\begin{tabular}{ccc}
\includegraphics[width=5.5cm,height=4cm]{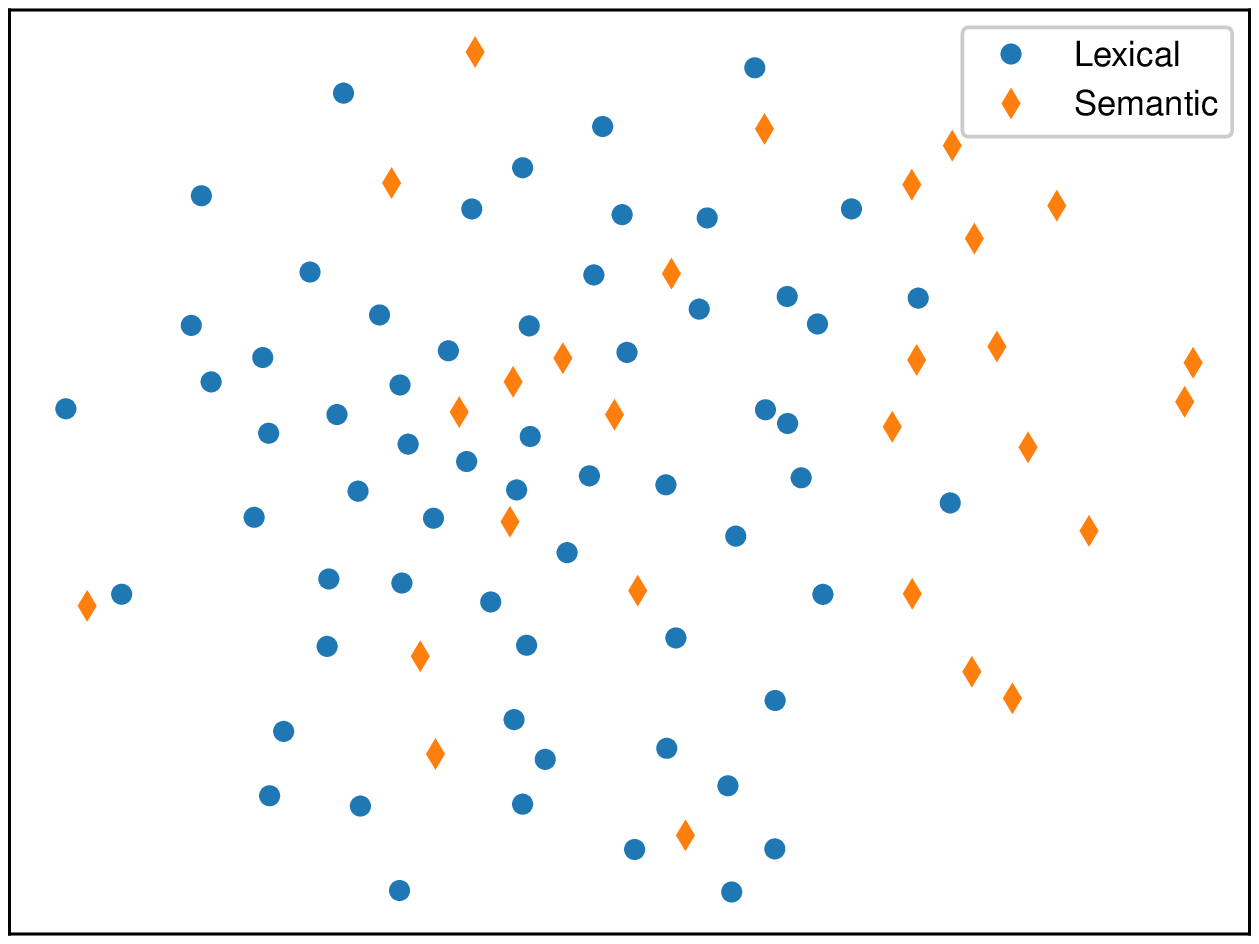} &
\includegraphics[width=5.5cm,height=4cm]{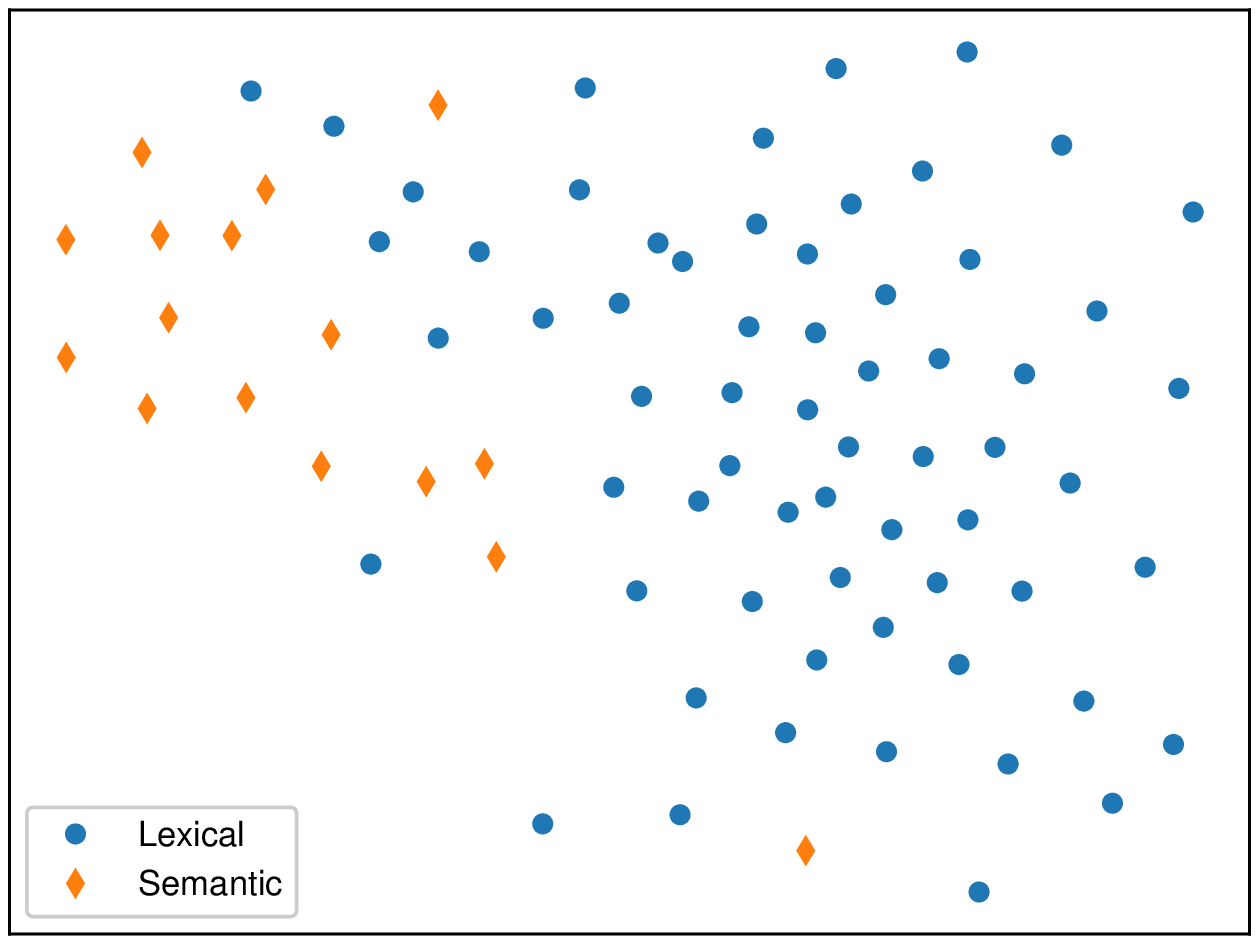} &
\includegraphics[width=5.5cm,height=4cm]{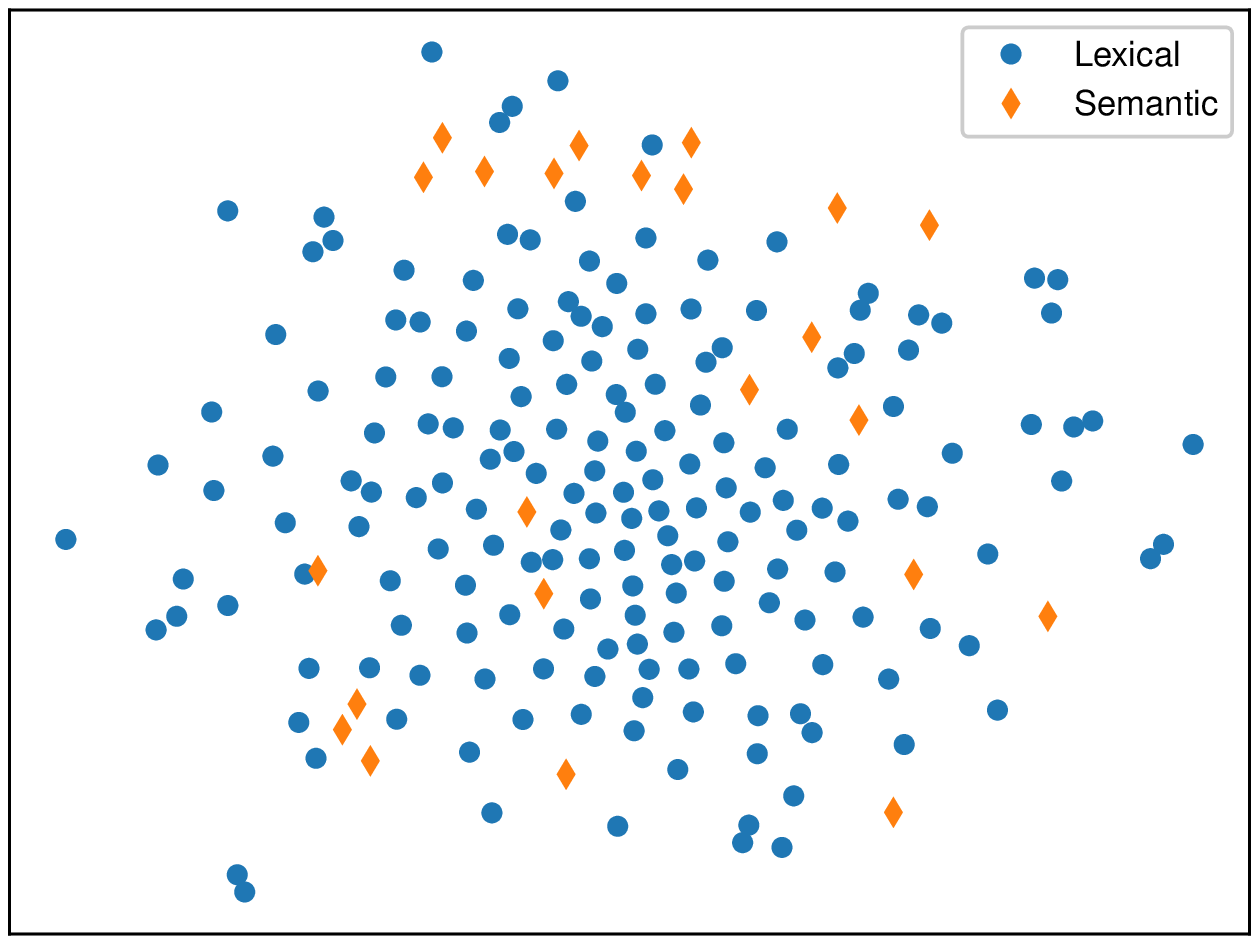} \\
(a) \textit{weather related fatalities} & (b) \textit{automation} & (c) \textit{efforts to enact gun control legislation}
\end{tabular}
\caption{\label{fig:clusters} Two-dimensional visualization of the relevant documents in the lexical and the semantic retrieval models.}
\end{figure*}

In the next analysis, we are interested in examining the performance of different groups of queries, divided based on their performance when using the lexical retrieval model. 
This analysis can help us better understand the origin of the average overall improvements of the hybrid approach over the lexical model. 
The results of this analysis are reported in Table \ref{tab:quarters}. To perform the analysis, we split the query set into four equal groups (Q1-4) based on similar performance when using the lexical approach (Q1 are the poorest-performing queries). According to the results, we can see that the improvements of the hybrid approach are much higher for the low quarters with an average improvement of $14\%$ for Q1. In the higher quarter (Q4), on the other hand, there is only a very slight improvement. 
To further understand the different properties of queries in the different performance groups, we perform an analysis of different query properties. Specifically, for each query, the mean, max, and standard deviation of the $\mathit{idf}$ values of its terms is computed; the number of query terms is also calculated. The average values of these measures in each query group are reported in Table \ref{tab:properies}.
According to the results, the mean and max of $\mathit{idf}$ values is higher for the query groups in which the hybrid approach is better performing (for example, comparing the performance of Q1 with that of Q4). 
A possible explanation for this is that lexical approaches can fail in cases where the query is dominated by a single term that has a high $\mathit{idf}$ value. This might be the case since lexical models often weigh the importance of query terms using a function of $\mathit{idf}$. An example of such a scenario was also given in Table \ref{tab:doc-example} in the introduction. In that example, we saw that the lexical retrieval model ``missed'' a relevant document that did not contain a word with a potentially high $\mathit{idf}$.
This observation is further supported by the standard deviation values of the different groups, also reported in Table \ref{tab:properies}. Finally, the results show that the queries with the better performance when using the hybrid approach are longer. A possible explanation for that is the ability of neural networks to learn semantics using multiple words.

\begin{table}[h]
	\center
    \caption{\label{tab:properies} Different properties of query groups, partitioned based on their performance when the lexical model was used.}
	\begin{tabular}{lrrrr}
\toprule
Property & Q1 & Q2 & Q3 & Q4 \\ 
\hline
$\mathit{Mean}(\mathit{idf})$ & $10.4$ & $10.4$ & $10.3$ & $9.3$ \\
$\mathit{Max}(\mathit{idf})$ & $16.9$	& $16.0$ & $16.4$ & $15.2$ \\
$\mathit{Std}(\mathit{idf})$ & $6.9$ & $6.0$ & $6.5$ & $5.9$  \\
Number of terms & $3.8$ & $3.9$ & $3.3$ & $3.7$ \\
\bottomrule
\end{tabular} 
\end{table}

\subsubsection{An analysis of relevant documents}
\begin{table}[t]
	\center
	\scriptsize
    \caption{\label{tab:example} Representative terms in relevant documents which were retrieved by the different retrieval models. Boldface: a unique term for a specific model.}
	\begin{tabular}{cc|cc|cc}
\toprule
\multicolumn{2}{c|}{\makecell{(a) \textit{weather related fatalities}\\ \\($\#docs=28$; $J=.333$)}} & \multicolumn{2}{c|}{\makecell{(b) \textit{automation}\\ \\($\#docs=16$; $J=.176$)}} & \multicolumn{2}{c}{\makecell{(c) \textit{efforts to enact gun}\\ \textit{control legislation}\\ ($\#docs=23$; $J=.282$)}} \\ 
\hline
Lexical & Semantic & Lexical & Semantic & Lexical & Semantic \\
\hline
people & storm & \textbf{automation}	& system & gun & gun\\
storm &	wind & system & \textbf{data} & \textbf{bill}	& \textbf{bush}\\
head & head	& \textbf{product} & \textbf{application}	& nra &	\textbf{text}\\
weather & \textbf{hurricane}	& \textbf{automate} & software & \textbf{control} & weapon\\
report & people & \textbf{operation} & \textbf{information} & \textbf{drug} & ban\\
\textbf{tornado} & \textbf{mph} & \textbf{center} & \textbf{new} & law & \textbf{say}\\
wind & weather & \textbf{process} & \textbf{service} & weapon	& \textbf{president}\\
\textbf{home} & \textbf{island} & \textbf{staff} & \textbf{user} & \textbf{handgun} & law\\
\textbf{today} & report & software & \textbf{image} &	ban & \textbf{issue}\\
\textbf{service} & \textbf{inch} & \textbf{management} & \textbf{ibm} & \textbf{wait} & nra\\
\bottomrule
\end{tabular}
 
\end{table}

In the following, an analysis is performed to shed light on the differences between the relevant documents retrieved by the lexical and the semantic models.

We start the analysis with a case study of three example queries from the query set. 
These queries were selected since they contain a substantial amount of relevant documents for the two retrieval models, cover diverse topics, and are of different lengths.
The first result of this analysis is presented in Table \ref{tab:example}. 
For each query, a semantic and a lexical list of 1000 documents is retrieved. Then, representative terms are extracted from each list using the top $k$ relevant documents in the list, where $k$ is set to be the minimum number of relevant documents between the two lists. The representative terms are then extracted using the $\mathit{tf.idf}$ scoring function. For each query, the number of relevant documents used, $\#docs$, and the Jaccard index ($J$) between the term lists of the two approaches (of 50 terms) are reported in the header line. 
Query (a) (``weather related fatalities'') is an example of the case where the semantic terms are related to a narrow topic, while the lexical terms cover a more general topic. Specifically, the semantic list has terms related to the topic of hurricanes (e.g., ``hurricane'', ``island'', and ``mph''), while the lexical terms are all related to the theme of the query, but can hardly be associated with a single topic. 
In such a case, the hybrid approach can potentially improve over the lexical baseline by strengthening the coverage of a specific aspect of the information need.
Query (b) (``automation'') is an example of a case in which the two approaches presumably cover two distinct topics. The semantic terms are quite related to the aspect of computer automation (e.g., ``ibm'' and ``application''), wherein the lexical retrieval model we can see terms related to automation in the traditional industry (e.g., ``product'' and ``staff'').
Query (c) (``efforts to enact gun control legislation'') serves as another example for a situation in which the semantic results presumably cover a narrow topic. Specifically, the terms ``president'' and ``bush'' might insinuate that.
Quantitatively, we can see that the vocabulary of the documents is substantially different for the two models as supported by the low Jaccard index.

\begin{table}[h]
	\center
    \caption{\label{tab:jaccard} The mean and standard deviation of the Jaccard index between the representative terms of the semantic and the lexical retrieval models for different number of terms.}
	\begin{tabular}{cccccc}
\toprule
\multicolumn{2}{c}{} & 10  & 50 & 100 & 200 \\
\hline

\multirow{2}{*}{Jaccard} & Mean & .184 &	.169&	.162 &	.156 \\
& Std & .153 &	.102	& .091 &	.094\\ 

\bottomrule
\end{tabular} 
\end{table}

The difference between the relevant documents of the two approaches is further emphasized by the visualization presented in Figure \ref{fig:clusters}. 
In the figure, the relevant documents of the two approaches are placed in a two-dimensional space using their $\mathit{tf.idf}$ representations.\footnote{The vocabulary was restricted to words that appear in at least 10 documents in each document set of a given query.} 
We focus only on documents that are unique for a specific retrieval model.
The vectors were embedded into a two-dimensional space using the t-SNE technique \cite{maaten2008visualizing};
According to the visualization, it can be seen that the semantic results often form clusters that are located in areas with small (or no) presence of lexical results. In some cases (query (c), for example), the lexical results can form a single dense cluster and the semantic results appear in sparser areas. This analysis shows the potential of the hybrid approach in increasing the diversity and the topic coverage of the result list.

To further support the above findings, a quantitative analysis was performed. 
For the analysis, all queries with at least five relevant documents, retrieved by each retrieval model, were taken into consideration, resulting in 50 queries. 
For each query, only the first five relevant documents were used to eliminate any biases regarding the number of documents considered. Then, we examined the average and the standard deviation of the Jaccard index between the term lists of the semantic and the lexical models; this analysis was performed for different numbers of terms. The results, presented in Table \ref{tab:jaccard}, show that, in the general case, the overlap between terms in the semantically retrieved documents and the lexically retrieved documents is very low. Moreover, this finding is consistent for different lengths of the term list and is stable over queries as can be attested by the low standard deviation.

\begin{figure}[t]
\center
\includegraphics[width=6.5cm,height=4.8cm]{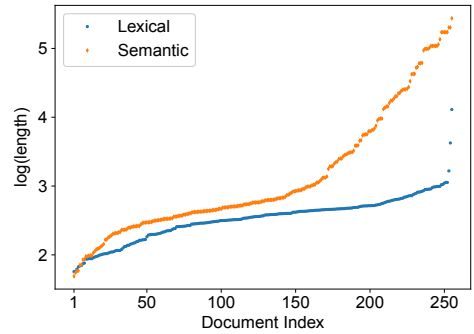}
\caption{\label{fig:lengths} The lengths of relevant documents, retrieved by the semantic and the lexical retrieval models.}
\end{figure}

The relevant documents in the two approaches can also differ in length as can be seen in Figure \ref{fig:lengths}. To construct the figure, the relevant documents with respect to all queries were pooled, sorted by length, and finally placed in a scatter plot.\footnote{We used $5$ documents per query, resulting in 250 documents overall; note that a point on the x-axis usually refers to two different documents.}
We can see from the figure that the semantic-based documents are often longer than the lexical-based documents and for about half of these documents the difference can be very large. 
A possible explanation for this is that classical lexical retrieval models are often designed to penalize long documents in the scoring function. This mechanism, however, does not exist in the semantic-based approaches. Furthermore, it might be the case where semantic approaches can better leverage longer pieces of text and words with low frequencies by using dense representations. Consequently, semantic approaches may be better in retrieving long relevant documents. 
\section{Conclusions}
\label{sec:conc}
Lexical-based retrieval models are the common models used in search engines for the retrieval stage. 
This work is the first one to systematically study the combination of semantic and lexical models for the retrieval stage of the ad hoc document retrieval task.
We proposed a general hybrid approach for document retrieval that leverages both semantic and lexical retrieval models.
An in-depth empirical analysis was performed which demonstrated the effectiveness of the hybrid approach and also shed some light on the complementary nature of the lexical and the semantic models.



There are several possible directions for future work that can be tackled. First is the development of more sophisticated approaches for the merging of the lexical and the semantic result lists. Second, in this work we addressed the problem of representing long documents through breaking them into short passages. Instead, more complex representations that take into account document structure can be considered. Finally, it would be interesting to evaluate the effectiveness of the hybrid retrieval approach for other information retrieval tasks including question answering, recommendation systems, and conversational agents.

\balance

\bibliographystyle{ACM-Reference-Format}
\bibliography{biblio.bib}
\end{document}